# Self-induced inverse spin Hall effect in ferromagnets: demonstration through non-monotonous temperature-dependence in permalloy

O. Gladii,[*,1] L. Frangou,[1] A. Hallal,[1] R. L. Seeger,[1] P. Noël,[1] G. Forestier,[1] S. Auffret,[1] M. Rubio-Roy,[1] P. Warin,[1] L. Vila,[1] S. Wimmer,[2] H. Ebert,[2] S. Gambarelli,[3] M. Chshiev,[1] and V. Baltz[**,1]

[1] *Univ. Grenoble Alpes, CNRS, CEA, Grenoble INP, SPINTEC, F-38000 Grenoble, France*

[2] *Universität München, Department Chemie, Butenandtstr. 5-13, D-81377 München, Germany*

[3] *Univ. Grenoble Alpes, CEA, SYMMES, F-38000 Grenoble, France*

[*] *olga_gladiy@ukr.net*

[**] *vincent.baltz@cea.fr*

**Abstract**

We investigated the self-induced inverse spin Hall effect in ferromagnets. Temperature (*T*), thickness (*t*) and angular-dependent measurements of transverse voltage in spin pumping experiments were performed with permalloy films. Results revealed non-monotonous *T*-dependence of the self-induced transverse voltage. Qualitative agreement was found with first-principle calculations unravelling the skew scattering, side-jump, and intrinsic contributions to the *T*-dependent spin Hall conductivity. Experimental data were similar whatever the material in contact with permalloy (oxides or metals), and revealed an increase of produced current with *t*, demonstrating a bulk origin of the effect.

The generation of a spin current and its further conversion to a charge current have attracted considerable attention, facilitating advances in basic physics along with the emergence of closely related applications in the field of spintronics [1,2]. The electronic transport regime considers spins carried by conduction electrons, in contrast to the magnonic transport regime which refers to excitation of localized-magnetic-moments [3,4]. Electronic spin current can be considered to occur through two distinct mechanisms: drift 'spin-polarized' current, when spins are carried by conduction electrons drifting due to the effect of an electric field; and diffusive 'pure' spin current, which is caused by diffusion of conduction electrons bearing majority spin and minority spin in opposite directions. Non-magnetic metals (NM) only permit diffusive spin current, but both types of current can coexist in ferromagnetic metals (F). In the case of diffusive spin current, diffusion results from non-equilibrium conditions creating a spin imbalance. This imbalance can be triggered by several mechanisms including distinct densities of states at the interface between materials of different types (e.g. F and NM), and transfer of angular momentum between phonons, photons, and electrons [2]. In this context, an electrical current can be converted to a spin current and vice versa as a result of the spin-orbit interaction (SOI), which links the spin and the orbital angular momentum of an electron. As a result of SOI, a flow of charges (spin) causes transverse spin (charge) to accumulate [5]. One of the related effects of this phenomenon, known as the inverse spin Hall effect (ISHE) [6,7], is commonly used to study SOI in NMs inserted into archetypal F/NM bilayers. In some of these studies, a spin current is pumped from the F spin-injector at resonance [8,9], and the ISHE ensures spin-charge conversion in the NM [10]. The contribution of the F to spin-charge conversion can be difficult to distinguish from that of the NM, and spin-charge conversion arising from the F is frequently neglected in experiments [11–17]. However, as we will further discuss below, in some cases spin-charge conversion in the F may prevail and need to be carefully considered. Some experimental studies indicated that self-induced charge current can be generated at room temperature (*T*) in NiFe [13], Co [15] and Fe [15] ferromagnets at resonance. The proposed mechanism for the origin of this spin current considered asymmetric spin-dependent scattering at the different interfaces. More specifically, when magnetic moments precess, the angular momentum of 3d-electrons is transferred to 4s-conduction electrons leading to a spin-polarized current in the F. Spins then flow in a diffuse manner due to non-uniform magnetization, which has been ascribed to asymmetric spin relaxation at the various interfaces. SOI in the F further ensures spin-charge conversion via the ISHE. Experimental data indicated a conversion efficiency of about 1% for NiFe [13].

In this study, we investigated the self-induced ISHE in single permalloy thin films when brought to resonance. Most importantly, our results demonstrated the bulk origin of the effect. Our experimental data revealed the self-induced transverse charge current to have a non-monotonous *T*-dependence. This finding was corroborated by the results of first-principle calculations describing the various contributions to the *T*-dependent spin Hall conductivity.

The full stacks used were (from substrate to surface): //Cu(6)/NiFe($t_{NiFe}$=8;12;16;24;32)/Cu(3)/Al(2)Ox (nm) multilayers. Stacks were deposited at room-*T* by dc-magnetron sputtering on Si/SiO$_2$(500)// substrates at a pressure of 2.3 x 10$^{-3}$ mbar under argon. The NiFe layer was deposited from a Ni$_{81}$Fe$_{19}$ (at. %) permalloy target. A 2-nm-thick Al cap was deposited to form a protective Al(2)Ox film after oxidation in air. The sample dimensions were: *l* = 2.46 mm and *w* = 0.46 mm. Both sides of the samples were connected to electrodes using aluminum-wire-bonding.

Spin pumping experiments (Fig. 1(a)) were conducted in a continuous-wave electron paramagnetic resonance spectrometer. The sample was fitted with a three-loop-two-gap resonator operating at 9.6 GHz. An input power of 40 mW was used, corresponding to a value of excitation magnetic field of about $h_{rf}$ ~0.5 Oe. The precise value was determined for each data point by measuring the quality factor of the cavity. The resonator is similar to an X-band rectangular cavity operating in TE$_{102}$ mode [18,19]. **h**$_{rf}$ was thus applied along the *y* direction. A dc bias field (**H**) was simultaneously applied at an angle ($\theta$) with respect to the sample normal (*z*). For each angle tested, the amplitude of **H** was scanned across the resonant condition for the NiFe layer's magnetization (**M**). The corresponding electric potential difference (*V*) induced along the *y* direction as a result of spin pumping and spin-charge conversion was then recorded. The field-sweep-rate was about 14 Oe.s$^{-1}$. A typical *V* vs *H* spectrum is shown in Fig. 1(b). The symmetric (*Sym*) and the antisymmetric (*Antisym*) contributions were disentangled by fitting data using the following equation: $V = V_{sym}\Delta H^2/[(\Delta H_{pp}\sqrt{3}/2)^2 + (H - Hres)^2] - V_{antisym}(\Delta H_{pp}\sqrt{3}/2)(H - H_{res})/[(\Delta H_{pp}\sqrt{3}/2)^2 + (H - Hres)^2]$, where $\Delta H_{pp}$ is the the peak-to-peak line width, and $H_{res}$ is the resonance field. $V_{sym}$ can be produced by the ISHE combined with any contributions from the anisotropic magnetoresistance (AMR) effect - planar Hall effect (PHE) part - and the anomalous Nernst effect (ANE). $V_{antisym}$ generally results from the anomalous Hall effect. In addition to these measurements, the electromagnetic signal reflected by the sample was converted into an electrical signal by a Schottky diode, thus allowing absorption spectra to be measured (Fig. 1(c)). Lock-in detection was used to enhance the signal-to-noise ratio. Data were fitted using a Lorentzian derivative to determine $\Delta H_{pp}$ and $H_{res}$.

The experiments and data analysis described above were conducted at *T* ranging between 50 and 300 K (Fig. 2). The key novel result of our article is that, for NiFe, *V*$_{sym}$ displays a non-monotonous *T*-dependence. From Fig. 2(a), *V*$_{sym}$ can be seen to flip sign upon reversal of *H*. This observation agrees with the time-reversal symmetry properties of the ISHE [6,7]. The PHE, which is odd in *H*, can be omitted. Figures 2(b) and (c) also show that the non-monotonous *T*-dependence of *V* is not related to *V*$_{antisym}$ nor to the possible PHE, as deduced from the AMR [20] evolution obtained separately for *H* = 1 kOe using standard 4-point electrical measurements [21]. The non-monotonous $T$-dependence of $V$ was also independent of $\Delta H_{pp}$ vs $T$, which was monotonous (Fig. 2(c)) [22,23]. The total Gilbert damping was determined using the following equation: $\alpha = (\Delta H_{pp} - \Delta H_{pp0})\sqrt{3}|\gamma|/(4\pi f)$. Inhomogeneous broadening ($\Delta H_{pp0}$) due to spatial variations in the magnetic properties could reasonably be neglected when making estimations at 9.6 GHz, since $T$-invariant values of just a few Oe were found using similar samples and a broadband setup (compared to linewidth of the order of 25-30 Oe). The gyromagnetic ratio was determined by fitting data related to the $f$-dependence of $H_{res}$ at 300 K, and a reasonable value of $\gamma = 18.8$ MHz.Oe$^{-1}$ was obtained. In line with [24], a potential $T$-dependent change in the direction of anisotropy could also be ruled out from the behavior of $H_{res}$ vs $T$ (Fig. 2(d)). Data were satisfactorily described using the usual Kittel formula [25].

To gain further insight into the origins of $V_{sym}$, we performed angular($\theta$)-dependent measurements for $T = 95$ K (maximal signal). Experimental data were compared to numerical calculations (Fig. 3(a)). The following set of equations describing equilibrium conditions was considered [10,20,26]: $2H_{res}\sin(\theta - \theta_M) + 4\pi M_S \sin(2\theta_M) = 0$; and $(\omega/\gamma)^2 = [H_{res}\cos(\theta - \theta_M) - 4\pi M_S \cos(2\theta)][H_{res}\cos(\theta - \theta_M) - 4\pi M_S \cos^2(\theta_M)]$, where $M_S$ is the saturation magnetization and $\theta_M$ is the tilt in $M$. Numerical minimization returned $M_S = 700$ emu.cm$^{-3}$ and $\gamma = 18.5$ MHz.Oe$^{-1}$. The expression $\Delta H_{pp} = (2/\sqrt{3})\alpha(\omega/\gamma)/\cos(\theta - \theta_M) + |dH_{res}/d\theta|\Delta\theta$ [20,26] was used to describe the data shown in Fig. 3(b). Numerical minimization returned $\alpha = 0.008$, and $\Delta\theta = 0.25°$. The $\theta$-dependence of $\theta_M$ was also determined from the calculations and is plotted in Fig. 3(c). The related transverse voltage resulting from the ISHE was calculated by applying the following theoretical expression [10]: $V_{sym}(norm.) = \sin(\theta_M)[4\pi M_S \gamma \sin^2(\theta_M) + \sqrt{(4\pi M_S \gamma \sin^2(\theta_M))^2 + 4\omega^2}]/[(4\pi M_S \gamma \sin^2(\theta_M))^2 + 4\omega^2]$. The correspondence between experimental data and theoretical predictions (Fig. 3(d)) indicates that the ISHE may be the main effect influencing the $T$-dependence observed.

We then compared the charge current deduced from our experimental data (Fig. 4(a)): $I_C = [V_{sym,\theta=-90°}-V_{sym,\theta=90°}]/(2R)$, where $R$ is the resistance of the slab, to first-principle calculations of spin Hall conductivity (Fig. 4(b)). When performing calculations, the thin film was considered a bulk material. For these calculations, the spin-polarized relativistic-Korringa Kohn Rostoker (SPR-KKR) code was used [27–29]. In this code, the linear response Kubo formalism was implemented in a fully relativistic multiple-scattering KKR Green function method. Thermal effects were modeled by considering electron scattering due to lattice vibration to be the dominant effect, because application of $H$ in the ferromagnetic resonance (FMR) experiments quenched spin fluctuations. Coherent-potential approximation was used. The $T$-dependence of transversal spin Hall conductivity ($\sigma_{xy,NiFe}^{z}$) is in satisfactory qualitative agreement with the experimental findings, showing a non-monotonous behavior with a minimum around T=100 K. To gain more insight into the origins of the effect observed, we further disentangled the skew scattering ($\sigma_{xy,NiFe}^{sk}$) and side-jump plus intrinsic ($\sigma_{xy,NiFe}^{sj+intr}$) contributions to $\sigma_{xy,NiFe}^{z}$, based on an approach using scaling behavior [6,7,30]. The following equation was considered: $\sigma_{xy,NiFe}^{z} = \sigma_{xy,NiFe}^{sk}+\sigma_{xy,NiFe}^{sj+intr} = \sigma_{xx,NiFe}S+\sigma_{xy,NiFe}^{sj+intr}$, where S is the skewness factor. For every $T$ tested, $\sigma_{xx,NiFe}$ was varied by changing the composition of the alloy over a range from $Ni_{85}Fe_{15}$ to $Ni_{70}Fe_{30}$. S was subsequently determined from plots of $\sigma_{xy,NiFe}^{z}$ vs $\sigma_{xx,NiFe}$. The two contributions, $\sigma_{xy,NiFe}^{sk} = \sigma_{xx,NiFe}S$ and $\sigma_{xy,NiFe}^{sj+intr}$, were then plotted (Fig. 4(c)) to determine the $Ni_{81}Fe_{19}$ composition. The non-monotonous $T$-dependence of $\sigma_{xy,NiFe}^{z}$ could clearly be ascribed to the fact that the skew scattering and the side-jump plus intrinsic contributions have opposite signs and similar amplitudes. These results can be phenomenologically understood in the light of the resonant scattering model that takes split impurity levels into consideration [31,32]. By inserting the scattering phase shift of Fe in Ni, returned by the SPR-KKR code, into the equations for spin Hall proposed in [31], we determined the ratio between the skew scattering and side-jump contributions to be around -1.2. The same trend of opposing signs and similar amplitudes was observed. This finding also seems to infer that the intrinsic contribution to the ISHE is negligible in permalloy.

Interestingly, similar sets of experimental $T$-dependences for $I_C$ were obtained whatever the material in contact with the permalloy: $SiO_2$, MgO, AlOx oxides, Cu, and Pt metals. This observation further confirms the bulk origin of the effect (supplemental material), and also demonstrates that our observations are not linked to the ANE [33–35]. This effect could also generate a transverse charge current due to SOI, and shares the same symmetry as the ISHE. It is known to result from a $T$-gradient building up when maximum power is absorbed by the F. Because the thermal conductivity of the oxides used in our experiments is of the order of $W.m^{-1}.K^{-1}$ compared to a few hundred for the metals, significant changes in the amplitude of ANE-related observations is expected. However, our observations were independent of the heat-sinking efficiency of the stack. These results were also corroborated by the fact that the signal observed was independent of the field-sweep rate (supplemental material) [35].

We will now comment on the direction of the self-induced current ($J_{S,self}$) (Fig. 5). To gain further insight into this matter, a reference layer of Pt was added to the stack, either as a buffer or as a capping layer. In this case, spin-charge conversions produced by ISHE in the Pt and NiFe layers contribute concurrently to the total experimentally probed $I_C$. The Pt layer has a positive spin Hall angle ($\Theta_{ISHE,Pt}$). For sufficiently thick layers, $V_{sym}$ generated in Pt relates to $\Theta_{ISHE,Pt} l_{sf,Pt}$ because $\Theta_{ISHE,Pt}$ is known to be mostly due to intrinsic contributions [36–38]. $\Theta_{ISHE,Pt} l_{sf,Pt}$ and $V_{sym}$ are therefore $T$-independent. Furthermore, $V_{sym}$ in Pt flips sign when the stacking order or field are reversed [6,7]. Given this fact, and considering the electrical connections in our setup, a buffer Pt layer pumps a spin current ($J_{S,Pt}$) toward the substrate and returns a negative (positive) value of $V_{sym}$ for a field angle $\theta=-90°$ (90°), resulting in a negative value of $I_C = [V_{sym,\theta=-90°} - V_{sym,\theta=90°}]/(2R)$. Conversely, when a capping Pt layer is included, a positive value of $I_C$ is returned. The NiFe layer also has a positive Hall angle [13]. The findings presented in Fig. 5 therefore indicate that, with regards to spin current direction, the NiFe layer behaves similarly to a buffer Pt layer, as it induces a negative $I_C$. In this scenario, spin- and subsequent charge-currents in the Pt and NiFe layers add up for the buffer Pt layer case, and subtract for the capping case (inset of Fig. 5). Similar to previous experiments [13], the spin current may be generated as a result of asymmetric spin-dependent scattering across the NiFe film, possibly due to non-homogeneous film properties across its thickness and to subsequent asymmetric spin relaxation at the various interfaces. From these data, at T~95 K, the self-induced conversion of the NiFe can be as efficient as that observed with Pt. We also note that although spin-charge conversion in NiFe is inefficient close to 300 K and only relates to ISHE in the Pt layer, self-induced spin diffusion still occurs. This effect creates asymmetry in the subsequent spin-charge conversion and may contribute to the observed difference in $I_C$ measured at 300 K due to the inversion of the Pt growth order. Inverting the growth order also modifies the electric properties of the Pt layer and interfaces. For example, we measured a resistivity of $\sigma_{xx,Pt}$=4x10$^6$ S.m$^{-1}$ for the capping layer case and of 5x10$^6$ S.m$^{-1}$ for the buffer layer, which correspond to reasonable $l_{sf,Pt}$ values (~ 3-4 nm) for the spin diffusion length [38]. We note that, if $J_{S,self}$ were omitted, $\Theta_{ISHE,Pt} l_{sf,Pt}$ at 300 K could be calculated using the following equation: $\Theta_{ISHE,Pt} l_{sf,Pt} = \frac{I_C}{h_{rf}^2} \frac{1}{w\tanh[t_{Pt}/(2l_{sf,Pt})]} \frac{8\pi\alpha^2}{2eg_r^{\uparrow\downarrow}\gamma^2} \frac{(4\pi M_S\gamma)^2+4\omega^2}{4\pi M_S\gamma+\sqrt{(4\pi M_S\gamma)^2+4\omega^2}}$, where the spin mixing conductance is calculated from: $g_r^{\uparrow\downarrow} = 2\sqrt{3}\pi M_S \gamma t_{NiFe} \Delta H^{pump}/(g\mu_B \omega)$, with $\Delta H^{pump} = (\Delta H_{pp,NiFe/Pt} - \Delta H_{pp,NiFe})$ for the capping Pt layer case and $\Delta H^{pump} = (\Delta H_{pp,Pt/NiFe} - \Delta H_{pp,NiFe})$ for the buffer layer [10]. Using the parameters measured separately, $M_S$ = 700 emu.cm$^{-3}$, $\gamma$ =18.5 MHz.Oe$^{-1}$, $\Delta H_{pp,NiFe/Pt}$ = 57 Oe, $\Delta H_{pp,Pt/NiFe}$ = 48 Oe, $\Delta H_{pp,NiFe}$ = 29 Oe, we determined $g_r^{\uparrow\downarrow} = 27$ $n$m$^{-2}$ and 18 nm$^{-2}$ for the capping and buffer Pt layer cases, respectively. The $\tanh[t_{Pt}/(2l_{sf,Pt})]$ can be approximated to 1. When further considering the values of $I_C/h_{rf}^2$ returned from the data in Fig. 5 at 300 K, we calculated $\Theta_{ISHE,Pt}l_{sf,Pt} = 0.23$ and 0.52 nm for the capping and buffer Pt layer cases, respectively. These data give the expected order of magnitude for Pt [38]. The discrepancy between the two values tends to confirm that $J_{S,self}$ cannot be neglected when determining $\Theta_{ISHE,Pt}$, in agreement with [12].

We finally considered how the effect observed was affected by the NiFe layer thickness. We found that the position of maximum conversion, $I_{C,95K}$, was thickness-independent (Fig. 6(a)). This observation is in agreement with the bulk origin of the effect, described in our discussion of Fig. 4. We further observed that the amplitude of $I_{C,95K}$ showed a similar thickness-dependence to $I_{C,300K}$ (Fig. 6(b)). The thickness-dependence of $I_C$ relates to $t^*/\alpha^2$, where $1/\alpha^2$ accounts for the spin pumping efficiency, and $t^*$ describes the thickness-dependence of the spin-charge conversion efficiency [10]. The former parameter was found to increase with thickness in a linear fashion. This behavior is due to the decreasing role played by interfaces, and the subsequent decrease of $\alpha$ for thick layers [39]. For the conversion efficiency, in this case, the spin-sink is also the NiFe spin current generator. Considering that the spin current is due to asymmetric spin relaxation at the various interfaces, we get a situation similar to the case of a spin-sink receiving the spin current from a third party and can thus consider that $t^* = l_{sf,NiFe}\tanh[t_{NiFe}/(2l_{sf,NiFe})]$ [10]. $l_{sf,NiFe}$ was estimated by combining our measurements of longitudinal conductivity in the following relation [40]: $l_{sf,NiFe} = 0.91\sigma_{xx,NiFe}x10^{-12}$. The values calculated for $l_{sf,NiFe}$ at T=100 K ranged between 2.9 nm for 8-nm-thick NiFe films to 5.3 nm for the 32-nm-thick film, in agreement with [41]. Plotting $t^*/\alpha^2$ vs $T$ (inset of Fig. 6(b)) revealed a nearly linear behavior, corroborating the results of the thickness-dependence of $I_C$.

In conclusion, the main contribution of this paper is that it presents systematic evidence of a self-induced ISHE in FMR experiments. Our findings were supported by distinct sets of $T$-

, thickness-, angular-, and stack-dependent experimental data encompassing the main features of the self-induced ISHE. The experimental findings were corroborated by first-principle calculations. Most importantly, similar amplitudes but opposite signs for the bulk skew scattering and the side-jump plus intrinsic contributions to the $T$-dependent spin Hall conductivity in permalloy could explain why the SOI-related transverse voltage was observed to display non-monotonous $T$-dependence. The findings from this study contribute to our understanding of a previously overlooked and incompletely understood effect. The results further indicate that self-induced conversion within the ferromagnet can be as efficient as that recorded with noble metals such as Pt, and thus needs to be carefully considered when investigating SO-related effects in materials destined for use in spintronics.

**Acknowledgments**

We acknowledge financial support from the French national research agency (ANR) [Grant Number ANR-15-CE24-0015-01] and KAUST [Grant Number OSR-2015-CRG4-2626]. We also thank O. Sipr for fruitful discussions and M. Gallagher-Gambarelli for critical reading of the manuscript.

**Figure captions**

Fig. 1. (a) Schematic representation of the experiment design. (b) Representative data showing $H$-dependence of $V$, as measured for a Si/SiO$_2$//Cu(6)/NiFe(8)/Cu(3)/Al(2)Ox (nm) stack at 95 K, when $\theta = +90°$. (c) Corresponding differential absorption spectra ($d\chi''/dH$ vs $H$). The lines in (b) and (c) were fitted to the data, see text.

Fig. 2. (a) $T$-dependence of the symmetric contribution, $V_{sym}$ to $V$, normalized by 'the microwave power' proportional to $h_{rf}^2$. Data measured for a Si/SiO$_2$//Cu(6)/NiFe(8)/Cu(3)/Al(2)Ox (nm) stack, when $\theta = +90°$ and $-90°$. (b) $T$-dependences of the antisymmetric contribution, $V_{antisym}$ plotted along with the AMR. (c,d) $T$-dependences of $\Delta H_{pp}$ (and corresponding $\alpha$) and $H_{res}$. The line was obtained using the Kittel model.

Fig. 3. $\theta$-dependences of (a) $H_{res}$, (b) $\Delta H_{pp}$, (c) the tilt in magnetization $\theta_M$, and (d) $V_{sym}$. Symbols: data measured at 95 K for a Si/SiO$_2$//Cu(6)/NiFe(8)/Cu(3)/Al(2)Ox (nm) stack. Solid lines were obtained using models described in the text.

Fig. 4. (a) $T$-dependence of the charge current ($I_C$) generated by spin-charge conversion in an Si/SiO$_2$//Cu(6)/NiFe(8)/Cu(3)/Al(2)Ox (nm) stack. Inset: $T$-dependence of the NiFe layer's longitudinal conductivity ($\sigma_{xx,\text{NiFe}}$) obtained independently using standard 4-point electrical measurements. (b) $T$-dependence of the spin Hall conductivity of bulk NiFe ($\sigma_{xy,NiFe}^z$) determined from first-principle calculations. Inset: $T$-dependence calculated for $\sigma_{xx,\text{NiFe}}$. (c) Skew scattering ($\sigma_{xy,NiFe}^{sk}$) and side jump plus intrinsic ($\sigma_{xy,NiFe}^{sj+intr}$) contributions to $\sigma_{xy,NiFe}^z$.

Fig. 5. (a) *T*-dependence of the charge current ($I_C$) generated in Si/SiO$_2$//Pt(10)/NiFe(8)/Cu(6)/Al(2)Ox (buffer Pt), Si/SiO$_2$//Cu(6)/NiFe(8)/Pt(10)/Al(2)Ox (capping Pt), and Si/SiO$_2$//Cu(6)/NiFe(8)/Cu(3)/Al(2)Ox (nm) stacks. (b) Schematic representations of the spin and charge currents in the stacks.

Fig. 6. (a) *T*-dependence of $I_C$ measured in Si/SiO$_2$//Cu(6)/NiFe($t_{NiFe}$=8;12;16;24;32)/Cu(3)/Al(2)Ox (nm) stacks. (b) NiFe thickness-dependence of $I_C$ measured at 95 and 300 K. Inset: corresponding thickness-dependences of $t^*/\alpha^2$.

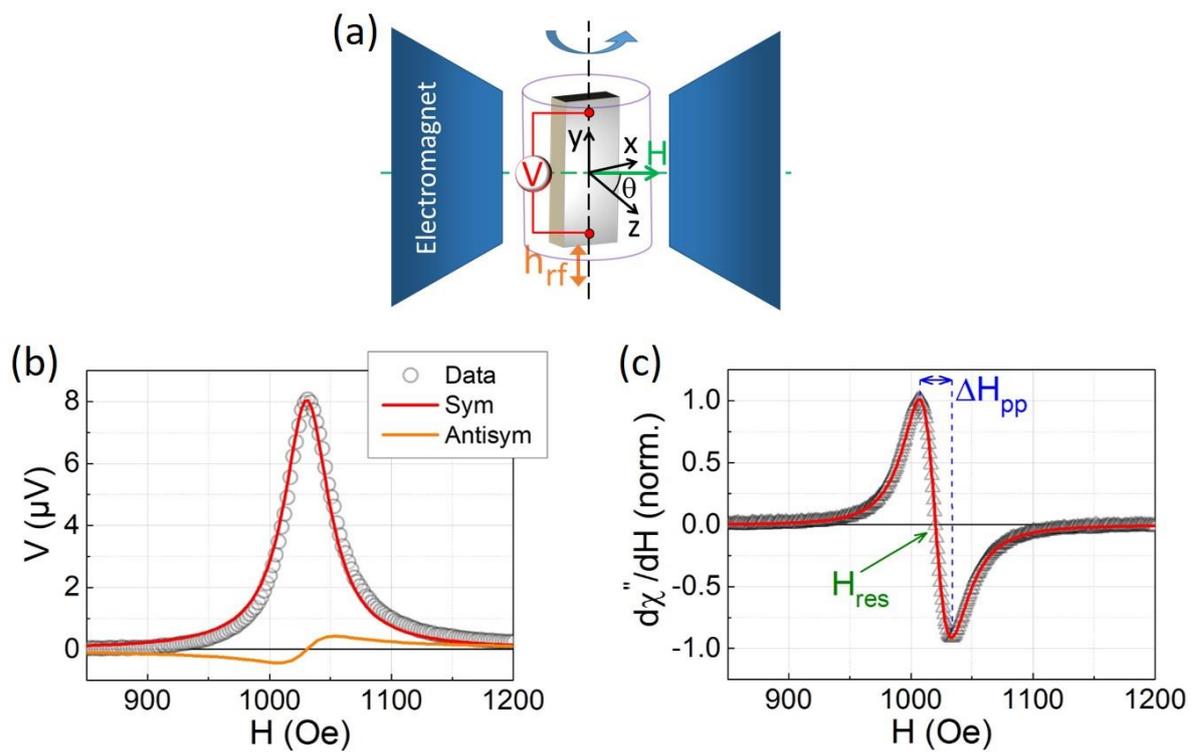

Fig. 1

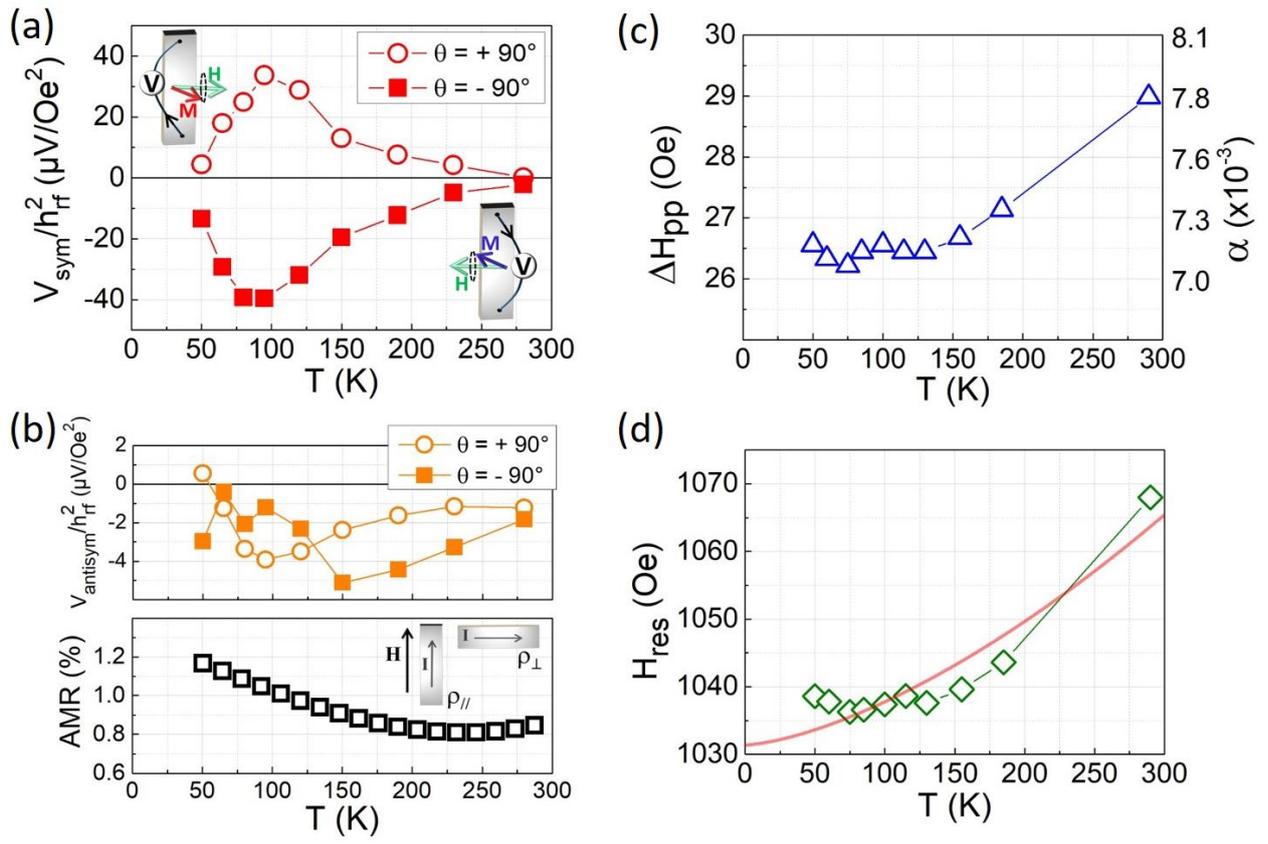

Fig. 2

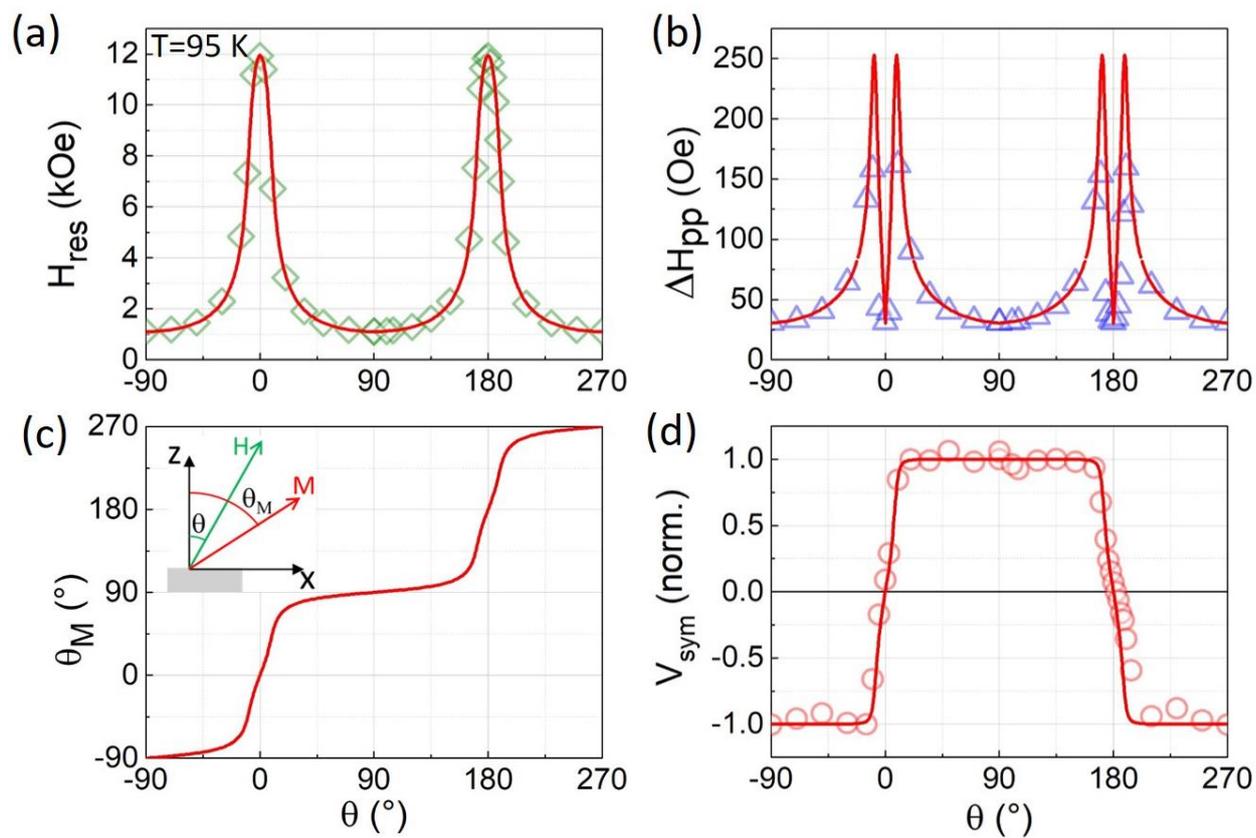

Fig. 3

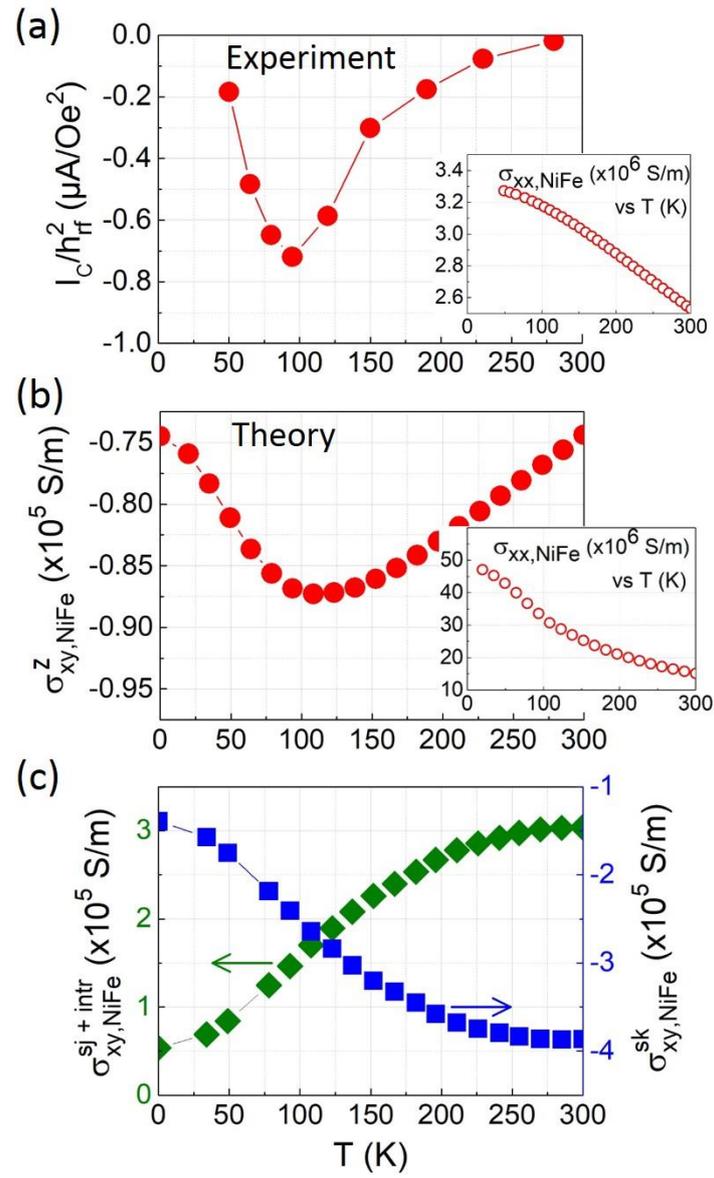

Fig. 4

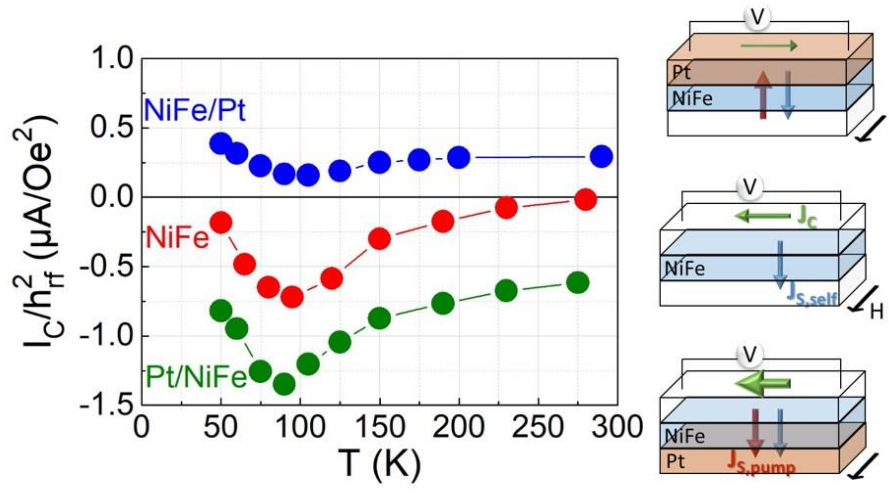

Fig. 5

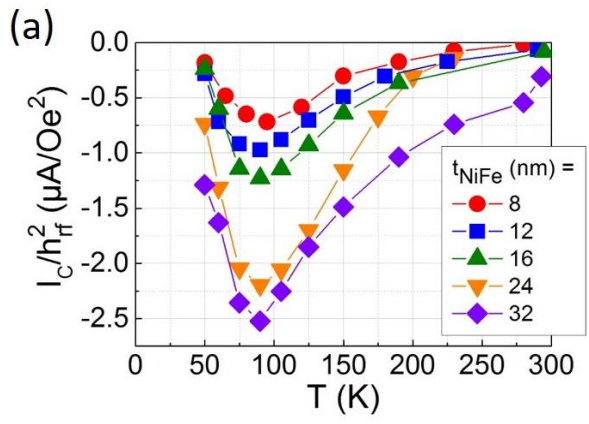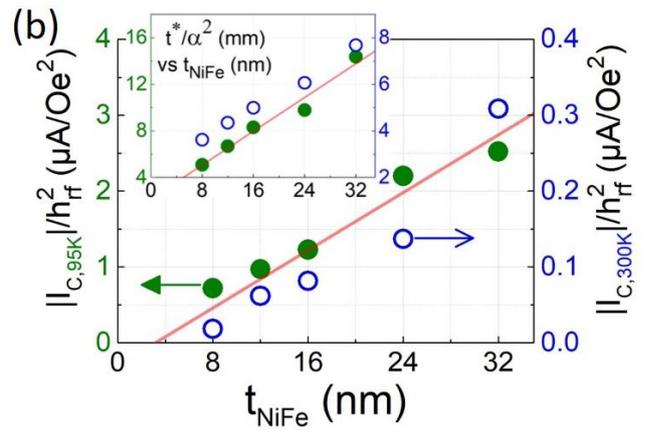

Fig. 6

# Supplemental material for 'Self-induced inverse spin Hall effect in ferromagnets: demonstration through non-monotonous temperature-dependence in permalloy'

O. Gladii,[*,1] L. Frangou,[1] A. Hallal,[1] R. L. Seeger,[1] P. Noël,[1] G. Forestier,[1] S. Auffret,[1] M. Rubio-Roy,[1] P. Warin,[1] L. Vila,[1] S. Wimmer,[2] H. Ebert,[2] S. Gambarelli,[3] M. Chshiev,[1] and V. Baltz[**,1]

[1] *Univ. Grenoble Alpes, CNRS, CEA, Grenoble INP, SPINTEC, F-38000 Grenoble, France*

[2] *Universität München, Department Chemie, Butenandtstr. 5-13, D-81377 München, Germany*

[3] *Univ. Grenoble Alpes, CEA, SYMMES, F-38000 Grenoble, France*

[*] *olga_gladiy@ukr.net*

[**] *vincent.baltz@cea.fr*

**1.** In this first part of supplemental material, we show that the non-monotonous *T*-dependence of spin-charge conversion was independent of the material in contact with the permalloy: $SiO_2$, MgO, AlOx oxides, Cu, and Pt metals, further confirming the 'bulk' origin of the effect. When not specified, samples were grown in the same machine, by sputtering.

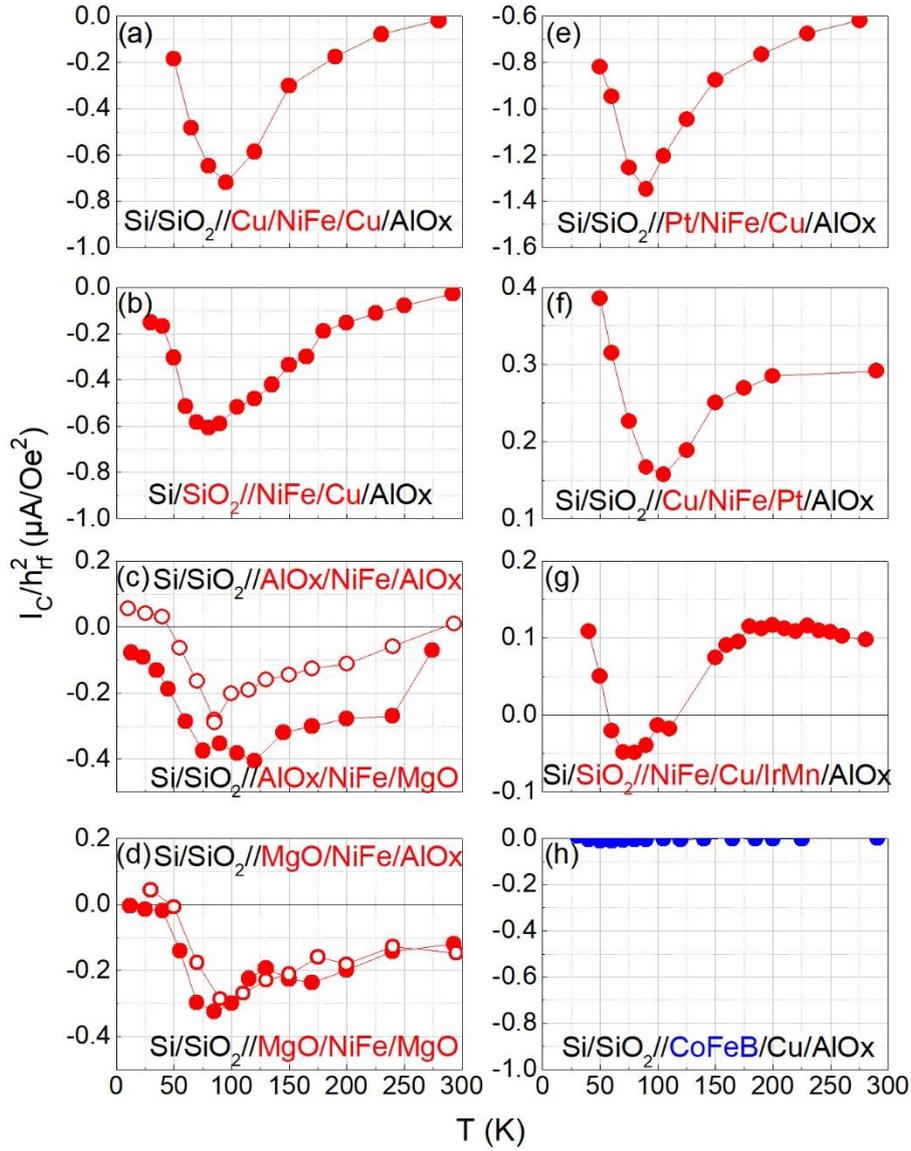

Fig. S1: (a) T-dependence of $I_C$ in several stacks. The compositions were: (a) $Si/SiO_2$//Cu(6)/NiFe(8)/Cu(3)/Al(2)Ox, (b) $Si/SiO_2$//NiFe(8)/Cu(3)/Al(2)Ox, (c) $Si/SiO_2$//AlOx(20)/NiFe(16)/AlOx(20) and $Si/SiO_2$//AlOx(20)/NiFe(16)/MgO(20), (d) $Si/SiO_2$//MgO(20)/NiFe(16)/AlOx(20) and $Si/SiO_2$//MgO(20)/NiFe(16)/MgO(20), (e) $Si/SiO_2$//Pt(10)/NiFe(8)/Cu(6)/Al(2)Ox, (f) $Si/SiO_2$//Cu(6)/NiFe(8)/Pt(10)/Al(2)Ox, (g) $Si/SiO_2$//NiFe(8)/Cu(3)/IrMn(1.5)/Al(2)Ox, and (h) $Si/SiO_2$//CoFeB(8)/Cu(3)/Al(2)Ox (nm).

Figure S1(a) is the reference sample, with the NiFe layer encapsulated between two metallic Cu layers.

The data in Figure S1(b) show that replacing a Cu/NiFe interface by an $SiO_2$/NiFe interface does not alter the temperature-dependence profile for the charge current.

For Figures S1(c) and (d), the NiFe/Cu and Cu/NiFe interfaces were replaced by interfaces with AlOx and MgO. These samples were grown in a different sputter machine. This difference explains the discrepancy in signal amplitude, and we cannot exclude the possibility that the quality of the NiFe stack grown on AlOx and MgO differ from that of the NiFe stack grown on $SiO_2$ or Cu. However, we note that the non-monotonous T-dependence of $I_C$ was nevertheless qualitatively similar for this set of samples.

Figure S1(e) corresponds to the Pt/NiFe/Cu stack – the buffer Pt case discussed in the main text. With this sample, spin-charge conversion in Pt shifts the signal downwards.

Figure S1(f) corresponds to the Cu/NiFe/Pt stack – the capping Pt case discussed in the main text. In this case, spin-charge conversion in the NiFe and Pt layers oppose one another. Replacing the Pt spin-charge converter by a Cu/IrMn layer induces a similar effect, confirming the findings. Note that for this latter case, Cu is used to avoid exchange bias coupling between the NiFe and IrMn layers.

Figure S1(g) shows that the effect was absent when the NiFe was replaced by a CoFeB layer.

**2.** In this second part of supplemental material, we present data that show that the non-monotonous *T*-dependence of spin-charge conversion is independent of the sweep rate used for the magnetic field (Fig. S2(a)), over the range accessible with our experimental setup. We note that this behavior is even valid despite an estimated temperature increase of up to about 180 mK, due to the absorption of the microwave power by the sample at resonance (Fig. S2(b-d)).

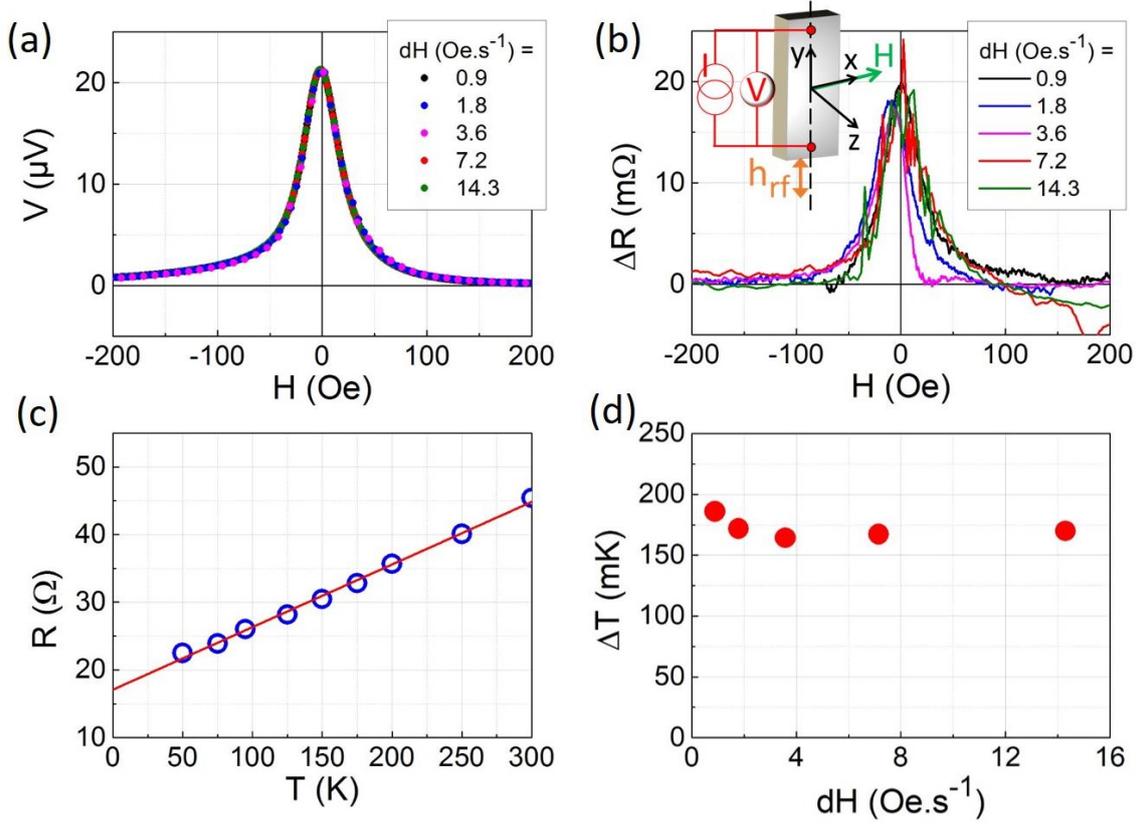

Fig. Sup. 2: (a) Representative *V* vs *H*-$H_{res}$ measured for a Si/SiO$_2$//Cu(6)/NiFe(32)/Cu(3)/Al(2)Ox (nm) stack at *T* = 95 K, when $\theta$ = + 90° (no bias current). Several sweep rates (*dH*) were used for the magnetic field. (b) The same dependences as in (a) were measured when a bias current of I = 100 µA was applied across the sample (see inset). The change in sample's resistance was estimated as follows, after removing the off-resonance voltage: $\Delta R$ = [$V_{with\ bias\ current}$ − $V_{without\ bias\ current}$]/I. (c) *T*-dependence of the off-resonance sample's resistance, measured independently. (d) *dH*-dependence of the increase in temperature of the sample at resonance ($\Delta T$), deduced from (b) and (c).